\documentclass{moriond}

\def\be{\begin{equation}}
\def\ee{\end{equation}}
\def\bea{\begin{eqnarray}}
\def\eea{\end{eqnarray}}

\newcommand{\gmn}{g_{\mu\nu}}
\newcommand{\beqn}{\begin{eqnarray}}
\newcommand{\eeqn}{\end{eqnarray}}
\newcommand{\nn}{\nonumber}
\newcommand{\dd}{\mathrm{d}}
\newcommand{\fmn}{f_{\mu\nu}}
\newcommand{\tr}{\mathrm{Tr}}

\newcommand{\Photo}{}

\begin{document}
\vspace*{4cm}
\title{CONSISTENT BIMETRIC THEORY AND ITS APPLICATION TO COSMOLOGY}

\author{ ANGNIS SCHMIDT-MAY }

\address{Department of Physics \& 
        The Oskar Klein Centre,\\
        Stockholm University, AlbaNova University Centre, 
        SE-106 91 Stockholm, Sweden}
\maketitle\abstracts{
Recently, the construction of ghost-free nonlinear massive spin-2 interactions solved a long outstanding problem in classical field theory. 
The consistent formulation requires a second rank-two tensor (or metric) and therefore has the form of a bimetric theory. Besides providing the solution to an old problem in field theory, ghost-free bimetric theory also has an interesting phenomenology: Its homogeneous and isotropic background solutions can reproduce the expansion history of the universe without any input of vacuum energy.}

\section{Introduction}

Our standard picture for cosmology, the $\Lambda$CDM model, relies on General Relativity (GR) as the theory for gravity and successfully describes the expansion history of the universe.
However, in this model, about 95$\%$ of the energy content in our universe is made of unknown constituents, commonly referred to as \textit{dark matter} and \textit{dark energy}. On top of that, we lack an explanation for the absence of a huge amount of vacuum energy (i.e.~a large cosmological constant) as expected from quantum field theory.

A lot of effort has been made to extend the Standard Model of Particle Physics and/or GR in order to solve the dark matter, dark energy and cosmological constant problems. 
Our approach here will not be to construct a new phenomenological model with the aim to account for the unexplained observational data. Instead, we address a fundamental question by constructing a model that is theoretically consistent. Once the consistent formulation is known, we  investigate its phenomenology and whether it can explain any of the observations. We will begin by stating a problem in classical field theory.
Massless and massive fields with spin 1 and lower play a fundamental role in the formulation of the Standard Model. Their consistent Lagrangians are known and it is possible to quantize them without introducing inconsistencies. For higher spins, however, the situation becomes more difficult. 

The unique \textit{classical} theory for a nonlinearly self-interacting massless spin-2 field is known to be GR formulated in terms of the metric tensor~$\gmn$, while its fundamental quantum version remains to be found.
The obvious next step in constructing classical field theories is to search for a Lagrangian describing massive spin-2 interactions. From our experience of fields with lower spin, we expect the massive theory to have the same kinetic structure as the massless one and we are therefore looking for a consistent deformation of the Einstein-Hilbert action of GR by a mass term for the metric~$\gmn$. This mass term should not contain any derivatives and, moreover, Lorentz invariance requires that it must not have any loose indices. But now we already face a problem because the only way to build a scalar out of $\gmn$ alone and without using derivatives is to contract its indices with the inverse metric $g^{\mu\nu}$. But since $g^{\mu\nu}\gmn=\tr\,(\gmn)=4$, this cannot give a nontrivial self-interaction term. In order to write down a mass term for a spin-2 field we therefore need to introduce a second rank-two tensor $\fmn$ that can be used to contract the indices on the metric. The interactions of the two fields will then be given in terms of traces of powers of the matrix $g^{\mu\nu}f_{\nu\rho}$. 

Clearly, the question arises whether the second metric $\fmn$ should be regarded as a fixed background or receive its own dynamics. From a field theory point of view, the latter choice seems more natural and we therefore also include an Einstein-Hilbert kinetic term for $\fmn$ into the action.
The result is a so-called bimetric theory of the form,
\beqn
S[g,f]=m_g^2\int\dd^4{x}\,\sqrt{g}\,R(g)+m_f^2\int\dd^4{x}\,\sqrt{f}\,R(f)-m^4\int\dd^4x\,\sqrt{g}~V(g^{-1}f)\,.
\eeqn
Here, the first two terms are standard Einstein-Hilbert kinetic terms for the two metrics with $m_f$ and $m_g$ setting their interaction strengths. The mass scale $m$ is related to the mass of the spin-2 field and $V$ is a generic interaction potential. Since the above action is invariant under diagonal diffeomorphisms, under which both $\gmn$ and $\fmn$ transform simultaneously, we expect the theory to contain a massless spin-2 excitation besides the massive mode. 

The problem with this kind of theory is that, for a general form of the interaction potential~$V$, it contains an additional scalar degree of freedom whose kinetic term has the wrong sign and therefore gives rise to a ghost instability.\cite{Boulware:1973my}
In the presence of this ghost, the Hamiltonian is unbounded from below which renders the theory unstable and hence inconsistent. In order to give a consistent description of massive spin-2 fields, it is therefore of uttermost importance to find a form of the interaction potential that eliminates the ghost mode.

\section{Ghost-free Bimetric Theory}

Although the linear theory for massive spin-2 in flat backgrounds had been known since 1939,\cite{Fierz:1939ix} its nonlinear version was constructed \cite{deRham:2010ik} \cite{deRham:2010kj} and proven to be consistent \cite{Hassan:2011hr} \cite{Hassan:2011tf} \cite{Hassan:2011zd} only very recently.
The unique interaction potential that avoids the ghost mode is of the form,\footnote{The corresponding massive gravity theory with nondynamical $\fmn$ is believed to give rise to superluminal propagation and acausality,\cite{Deser:2013eua} but so far these pathologies have not been found in bimetric theory.}
\beqn\label{intpot}
V(g^{-1}f)=\sum_{n=0}^4\beta_n e_n(S)\,.
\eeqn
Here, $\beta_n$ are arbitrary dimensionless parameters and $S\equiv\sqrt{g^{-1}f}$ is a matrix square root defined through $S^2=g^{-1}f$. Furthermore, $e_n(S)$ are the elementary symmetric polynomials of the matrix $S$ which can be expressed through a recursion formula,
\beqn
e_n(S)=\frac{1}{n}\sum_{k=0}^{n-1}(-1)^{n+k+1}\,\tr(S^{n-k})\,e_k(S)\,,\qquad
e_0(S)=1\,.
\eeqn
The appearance of the square root and the structure of the elementary symmetric polynomials are crucial for the absence of the dangerous ghost mode. The above potential with its five free parameters exhausts all possible ghost-free interaction terms for the two metrics. Note that, since $e_0(S)=1$ and $e_4(S)=\det S=(\sqrt{g})^{-1}\sqrt{f}$, the parameters $\beta_0$ and $\beta_4$ simply correspond to cosmological constant terms for $\gmn$ and $\fmn$, respectively.

Coupling the metrics to matter in the standard way (i.e.~as in General Relativity) does not reintroduce the ghost. In the simplest case, only one of the metrics couples to matter and thereby fixes the geometry and causal structure of the matter sector, while the other metric interacts with matter only indirectly.\footnote{The analysis of cosmological solutions has also been carried out with both metrics coupled to matter.\cite{Akrami:2013ffa}} Hence, with $\phi$ collectively denoting all the matter fields, the bimetric action that we will consider in the following is of the form,
\beqn\label{bimact}
S[g,f,\phi]=m_g^2\int\dd^4{x}\,\sqrt{g}\,R(g)&+&m_f^2\int\dd^4{x}\,\sqrt{f}\,R(f)\nn\\
&-&m^4\int\dd^4x\,\sqrt{g}~\sum_{n=0}^4\beta_n e_n(S)+\int\dd^4{x}\,\sqrt{g}\,\mathcal{L}_\mathrm{m}(g,\phi)\,.
\eeqn
The equations of motion obtained from the bimetric action~(\ref{bimact}) with interaction potential~(\ref{intpot}) by varying with respect to $\gmn$ and $\fmn$ read as,
\beqn\label{eom}
\mathcal{G}_{\mu\nu}(g)+\frac{m^4}{m_g^2}\mathcal{V}_{\mu\nu}(g,f)=\frac{1}{m_g^2}T_{\mu\nu}\,,\qquad
\tilde{\mathcal{G}}_{\mu\nu}(f)+\frac{m^4}{m_f^2}\tilde{\mathcal{V}}_{\mu\nu}(g,f)=0\,,
\eeqn
where $\mathcal{G}_{\mu\nu}(g)$ and $\tilde{\mathcal{G}}_{\mu\nu}(f)$ are the Einstein tensors for $\gmn$ and $\fmn$, respectively, $\mathcal{V}_{\mu\nu}(g,f)$ and $\tilde{\mathcal{V}}_{\mu\nu}(g,f)$ are the contributions from the interaction potential and $T_{\mu\nu}$ is the stress-energy tensor of the matter Lagrangian $\mathcal{L}_\mathrm{m}(g,\phi)$.

\section{Bimetric Cosmology}
In order to derive cosmological solutions in the theory specified by (\ref{bimact}), we make a homogeneous and isotropic ansatz for both of the metrics, such that $\gmn\dd x^\mu\dd x^\nu=-\dd t^2+a^2\dd x^2$ has the same form as in GR.\footnote{For simplicity, we assume vanishing curvature, $k=0$. The generalization to $k\neq 0$ is straightforward.} Furthermore, we assume the stress-energy tensor of the matter source to resemble a perfect fluid and be covariantly conserved, i.e.~$\nabla^\mu T_{\mu\nu}=0$, where $\nabla^\mu$ is the covariant derivative compatible with $\gmn$.
After a subset of the equations of motion (\ref{eom}) has been solved to determine the time-dependent functions in the ansatz for $\fmn$, the remaining equations are modified Friedmann equations for the scale factor $a(t)$ in the physical metric~$\gmn$,\cite{Volkov:2011an} \cite{vonStrauss:2011mq} \cite{Comelli:2011zm}
\beqn\label{modf}
\left(\frac{\dot{a}}{a}\right)^2=F_1[\rho]\,, \qquad 
\frac{\ddot{a}}{a}=F_2[\rho,p]\,,
\eeqn
where $F_{1,2}$ are functions (typically involving square, cubic or even quartic roots) of the energy density $\rho$ and the pressure $p$ of the matter source. Their precise form depends on the choice of interaction parameters. Generically, the evolution equations~(\ref{modf}) are very different from GR for which the functions $F_{1,2}$ are linear in $\rho$ and $p$. Hence, bimetric theory can give rise to very different dynamics and evolution of the universe. 

A particularly interesting question is whether it is possible to obtain an accelerating solution ($\ddot{a}>0$) with a matter source that contains relativistic and non-relativistic matter, but no vacuum energy. In GR this is impossible because, for sources with equation of state $w=p/\rho>-1/3$, acceleration cannot occur. In order to set the vacuum energy contribution in bimetric theory to zero, one assumes a source with only relativistic ($w=1/3$) and non-relativistic ($w=0$) matter components. Furthermore, one sets the cosmological constant term in the bimetric interaction potential to zero by fixing $\beta_0=0$. 

A detailed study of this setup by Akrami \textit{et.al.}~\cite{Akrami:2012vf} revealed that bimetric theory can indeed give rise to cosmological acceleration without the input of vacuum energy. Furthermore, in a statistical analysis the authors found that certain bimetric models can fit observational data and account for the expansion history of the universe just as well as standard $\Lambda$CDM. In these cases, the present Hubble scale is set by the mass parameter $m$ in the bimetric interaction potential.
\vspace{5pt}

It is expected that, even in cases where the background behaves very similarly to $\Lambda$CDM, bimetric theory will differ from GR at the level of perturbations around the homogeneous and isotropic backgrounds.\cite{Berg:2012kn} 
Progress in deriving predictions for CMB and structure formation has only been made very recently, indicating that, while compatible with observational data, bimetric theory will be distinguishable from GR in the near future.\cite{Konnig:2014dna} \cite{Solomon:2014dua} 
On the other hand, although the cosmological solutions are well-behaved for a large spin-2 mass,\cite{DeFelice:2014nja} a mass on the order of the present Hubble scale (as required for self-acceleration) could give rise to an instability.\cite{Comelli:2012db} \cite{Comelli:2014bqa}

\section{Summary $\&$ Discussion}

The construction of the consistent description for nonlinear spin-2 interactions in classical field theory resulted in a novel theory of modified gravity which contains two metrics interacting with each other. The cosmological background solutions in ghost-free bimetric theory are still compatible with observational data after all vacuum energy contributions have been removed. Hence, there is no need for a cosmological constant if GR is replaced by bimetric theory.

These results are promising due to the following reason: If a (yet unknown) mechanism or symmetry was able to set the vacuum energy contribution to zero at the quantum level, then the expansion of the universe could still accelerate due to the spin-2 interactions. 
In this case, the Hubble scale is set by the spin-2 mass which, unlike a cosmological constant, is protected against large quantum corrections due to the restoration of the full diffeomorphism invariance (under which both metrics transform independently) in the limit~$m\rightarrow 0$.

In order to be able to tell whether bimetric theory is really compatible with all cosmological observations, we still need a better understanding of its perturbation theory. In addition to this, it is necessary to compare the local solutions of the theory to observations in the solar system and verify that the same parameters can fit the data on all distance scales.

\end{document}